\documentclass[]{raa}
\usepackage{amssymb}            
\usepackage{graphicx,times}
\usepackage{natbib}
\usepackage{subfigure}
\providecommand{\bm}[1]{\mbox{\boldmath$#1$\unboldmath}}

\begin{document}

\title{Chromospheric activity on late-type star of V1355 Ori using Lijiang 1.8-m and 2.4-m telescopes
}

 \volnopage{ {\bf 20116} Vol.\ {\bf ?} No. {\bf XX}, 000--000}
   \setcounter{page}{1}

   \author{Qingfeng Pi
      \inst{1,2,3}
         \and Liyun Zhang
      \inst{1,2}
   \and Liang Chang
      \inst{2,4}
               \and Xianming L. Han
      \inst{5}
               \and Hongpeng Lu
      \inst{1,2}
         \and XiLiang Zhang
      \inst{2,4}
   \and Daimei Wang
      \inst{1,2}
   }

   \institute {College of Physics/Department of Physics and Astronomy, Guizhou University,
Guiyang 550025, P.R. China {\it piqingfeng@126.com; \it liy\_zhang@hotmail.com}\\
\and
Key Laboratory for the Structure and Evolution of Celestial Objects, Chinese Academy of Sciences, Kunming 650011 P.R. China
\and
MingDe College, Guizhou University, Guiyang 550025, P.R. China
\and
Yunnan Observatories, Chinese Academy of Sciences, Kunming, 650011, P.R. China
\and
Dept. of Physics and Astronomy, Butler University, Indianapolis, IN 46208, USA\\
\vs \no
   {\small Received ?; accepted ?}
}

\abstract{We obtained new high-resolution spectra using the Lijiang 1.8-m and 2.4-m telescopes to discuss the chromospheric activities of V1355 Ori as indicated in the behaviour in the $\mbox{Ca~{\sc ii}}$ H\&K, H$_{\delta}$, H$_{\gamma}$, H$_{\beta}$, $\mbox{Na~{\sc i}}$\ D$_{1}$ D$_{2}$, H$_{\alpha}$ and $\mbox{Ca~{\sc ii}}$ infrared triplet (IRT) lines. The observed spectra show obvious emissions above continuum in the $\mbox{Ca~{\sc ii}}$ H\&K lines, absorptions for the H$_{\delta}$, H$_{\gamma}$, H$_{\beta}$, and $\mbox{Na~{\sc i}}$\ D$_{1}$, D$_{2}$ lines, variable behaviour (filled-in absorption, partial emission with a core absorption component, or emission above continuum) for the H$_{\alpha}$ line, and weak self-reversal emissions in the strong filled-in absorptions of the $\mbox{Ca~{\sc ii}}$ IRT lines. We used a spectral subtraction technique to analyse our data. The results show no excess emission in the H$_{\delta}$ and H$_{\gamma}$ lines, very weak excess emissions for the $\mbox{Na~{\sc i}}$\ D$_{1}$, D$_{2}$ lines, excess emission in the H$_{\beta}$ line, clear excess emission for the H$_{\alpha}$ line, and excess emissions for the $\mbox{Ca~{\sc ii}}$ IRT lines. The value of the ratio of $EW_{8542}$/$EW_{8498}$ is in the range of 0.9 to 1.7, and it implies that the chromospheric activity might have been caused by plage events. The value of the ratio $E_{H_{\alpha}}$/$E_{H_{\beta}}$ is above 3, and it implies that the Balmer lines would arise from prominence-like material. We also found time variations of light curves of equivalent widths (EWs) of the chromospheric activity lines in the $\mbox{Na~{\sc i}}$\ D$_{1}$ D$_{2}$, $\mbox{Ca~{\sc ii}}$ IRT, and H$_{\alpha}$ lines in particular. These phenomena can be explained by plage events, which are consistent with the behaviour of chromospheric active indicators.  \\
\keywords{stars: chromosphere -- stars: activity -- stars: individual V1355 Ori}}
\authorrunning{Qingfeng Pi, Liyun Zhang, Liang Chang et al. }            
\titlerunning{Chromospheric activity of V1355 Ori}  
\maketitle
%
%
\section{Introduction}           
\label{sect:intro}
\indent V1355 Ori (BD -00 1147, HD 291095, K0-2 IV, 3.82 days) is a single-line spectroscopic RS CVn binary (Strassmeier et al. 1999; Strassmeier 2000; Strassmeier et al. 2000; Savanov \& Strassmeier 2008; Eker et al. 2008; etc). It exhibits photospheric and chromospheric magnetic activities, which has been discovered at optical (Cutispoto et al. 1995; Strassmeier 2000; Savanov \& Strassmeier 2008; etc), ultraviolet (Pounds et al. 1993), and X-ray wavelengths (Voges et al. 1999).\\
\indent Strassmeier et al. (2000) discovered that V1355 Ori exhibited very strong emission over the local continuum in $\mbox{Ca~{\sc ii}}$ H\&K lines, and showed weak emission in one single H$_{\alpha}$ spectra. Strassmeier (2000) detected a strong flare in the H$_{\alpha}$ line during doppler imaging observation in 1998. He also found that the H$_{\alpha}$ profile varies against the orbital phase during the three observing epochs in 1997, 1998, and 1999 (Strassmeier 2000). The profile of the H$_{\alpha}$ line appear to be quite complex, ranging from an extremely strong emission due to flaring emission with a central sharp absorption, and to partial emission with a red absorption component, and to very narrow absorption (Strassmeier et al. 2000; strassmeier 2000). No analysis has been carried out regarding the H$_{\delta}$, H$_{\gamma}$, H$_{\beta}$, $\mbox{Na~{\sc i}}$\, D$_{1}$, D$_{2}$, and $\mbox{Ca~{\sc ii}}$ IRT lines to date.\\
\begin{table*}
\tiny\tabcolsep 0.05cm \caption{Observation log of V1355 Ori using 1.8-m and 2.4-m telescopes.}
\begin{tabular}{cllcccccccccccccc}
\hline \hline\multicolumn{1}{c}{Date} &
\multicolumn{1}{c}{HJD(245,)} & \multicolumn{1}{c}{Exp.
time} & \multicolumn{10}{c}{$S/N$} \\
\cline{4-14} &days & (s)
& Ca II IRT$\lambda$8662&Ca II IRT$\lambda$8542& Ca II IRT$\lambda$8498&H$_{\alpha}$&Metal line
& $\mbox{Na~{\sc i}}$& H$_{\beta}$ (4861 {\AA})& H$_{\gamma}$ (4341 {\AA})
& H$_{\delta}$ (4102{\AA})& $\mbox{Ca~{\sc ii}}$ H (3968{\AA}) & $\mbox{Ca~{\sc ii}}$ K (3933{\AA}) \\
\hline
20150122   &   7045.22327   & 3600     &  58    &  62   &   52   & 52  &   81 &     70&        &         &     &     &    \\
20150124   &   7046.23188   & 1800     &  47    &  51   &   44   & 44  &   53 &     51&        &         &     &     &    \\
20150127   &   7050.17989   & 2400     &  45    &  50   &   43   & 43  &   54 &     52&        &         &     &     &    \\
20150204   &   7058.14409   & 3600     &  68    &  73   &   64   & 64  &   77 &     73&        &         &     &     &    \\
20150206   &   7060.11388   & 3600     &  67    &  71   &   62   & 62  &   81 &     75&        &         &     &     &    \\
20150209   &   7063.11839   & 1800     &  45    &  49   &   43   & 43  &   52 &     47&        &         &     &     &    \\
20150210   &   7064.13065   & 2400     &  49    &  53   &   45   & 45  &   58 &     55&        &         &     &     &    \\
20150211   &   7065.09594   & 3600     &  53    &  56   &   48   & 48  &   60 &     55&        &         &     &     &    \\
20150212   &   7066.11174   & 3600     &  57    &  63   &   54   & 54  &   68 &     65&        &         &     &     &    \\
20150213   &   7067.12605   & 3600     &  53    &  57   &   49   & 49  &   62 &     57&        &         &     &     &    \\
20150214   &   7068.12689   & 3600     &  58    &  67   &   55   & 55  &   76 &     71&        &         &     &     &    \\
20150215   &   7069.07940   & 3600     &  63    &  66   &   54   & 54  &   66 &     61&        &         &     &     &    \\
\hline
20160129   &   7417.15243   & 3600     &   87   &	84   & 	63    &67   & 89   &   74  &     28 & 	  17  &  	10&  	7 & 	6\\
20160226   &   7445.11321   & 3600     &   84   &	81   & 	61    &60   & 85   &   71  &     29 & 	  17  &  	10&  	6 & 	6\\
20160227   &   7446.07349   & 5400     &  151   & 145   &  110   &107  &	151  &   125 &     48 &	    30  &	16  &10   &  10\\

\hline
\end{tabular}
\end{table*}

\begin{figure*}
\includegraphics[width=6.8cm,height=7.8cm]{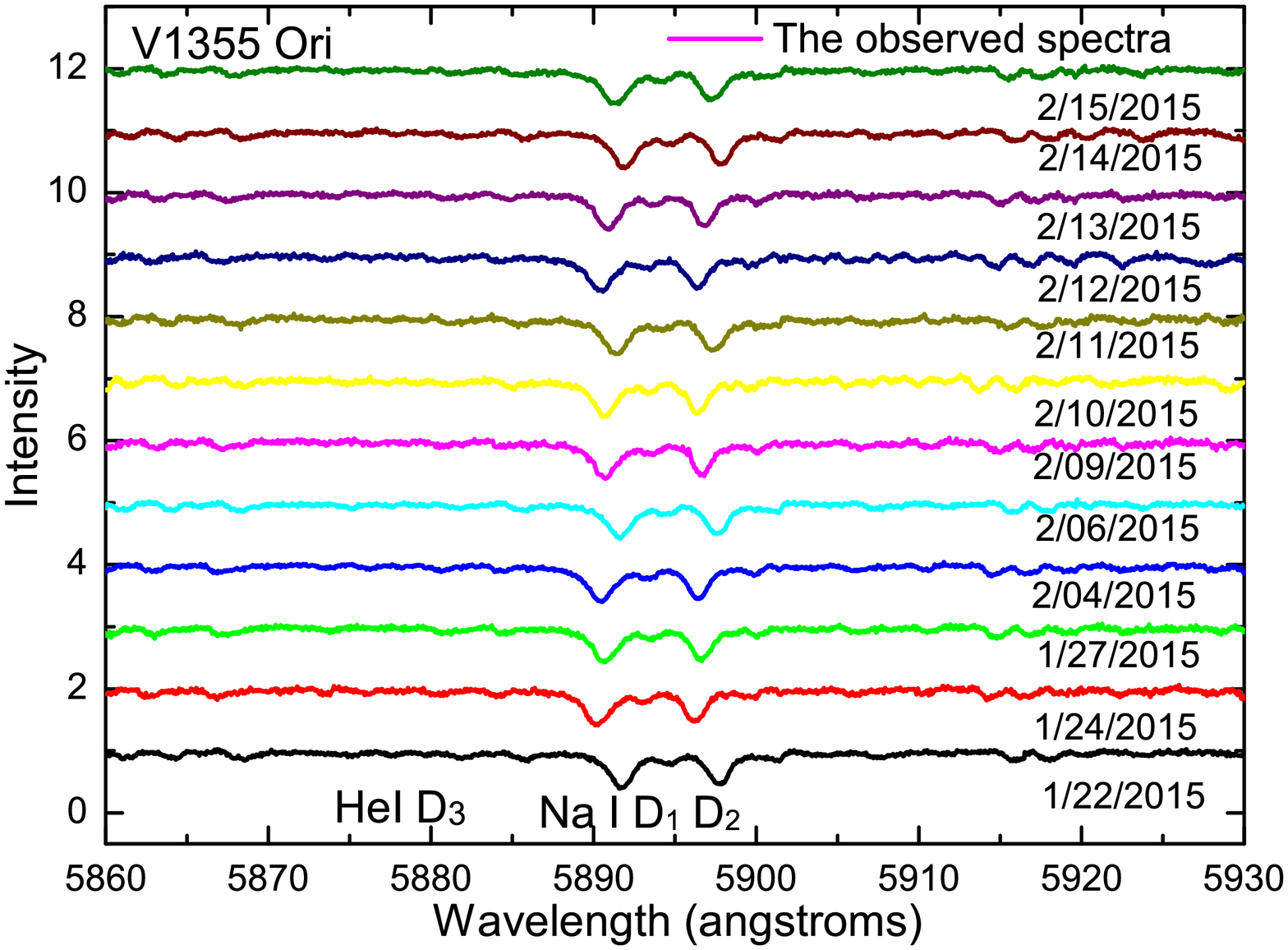}
\includegraphics[width=6.8cm,height=7.8cm]{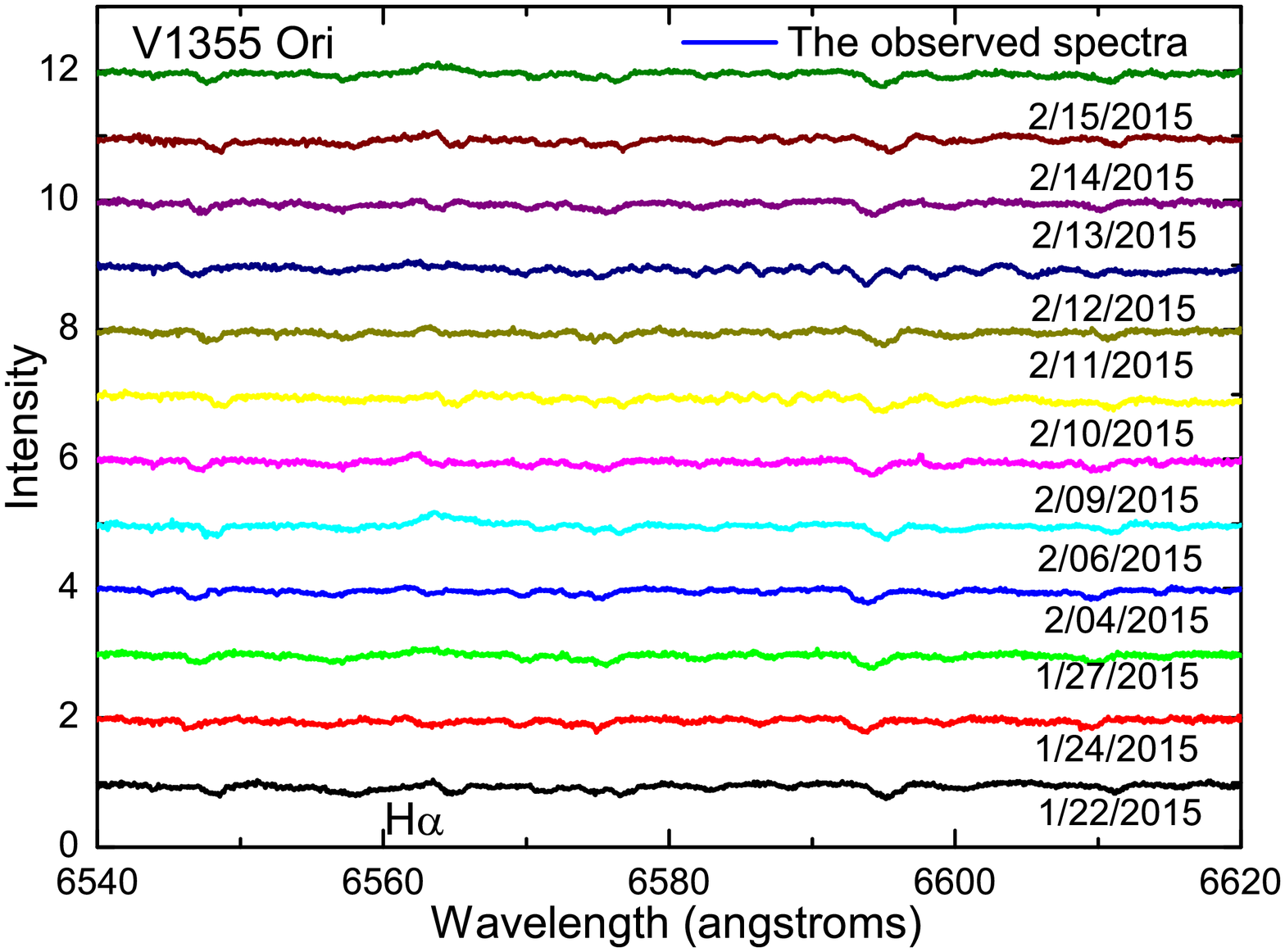}
\includegraphics[width=6.8cm,height=7.8cm]{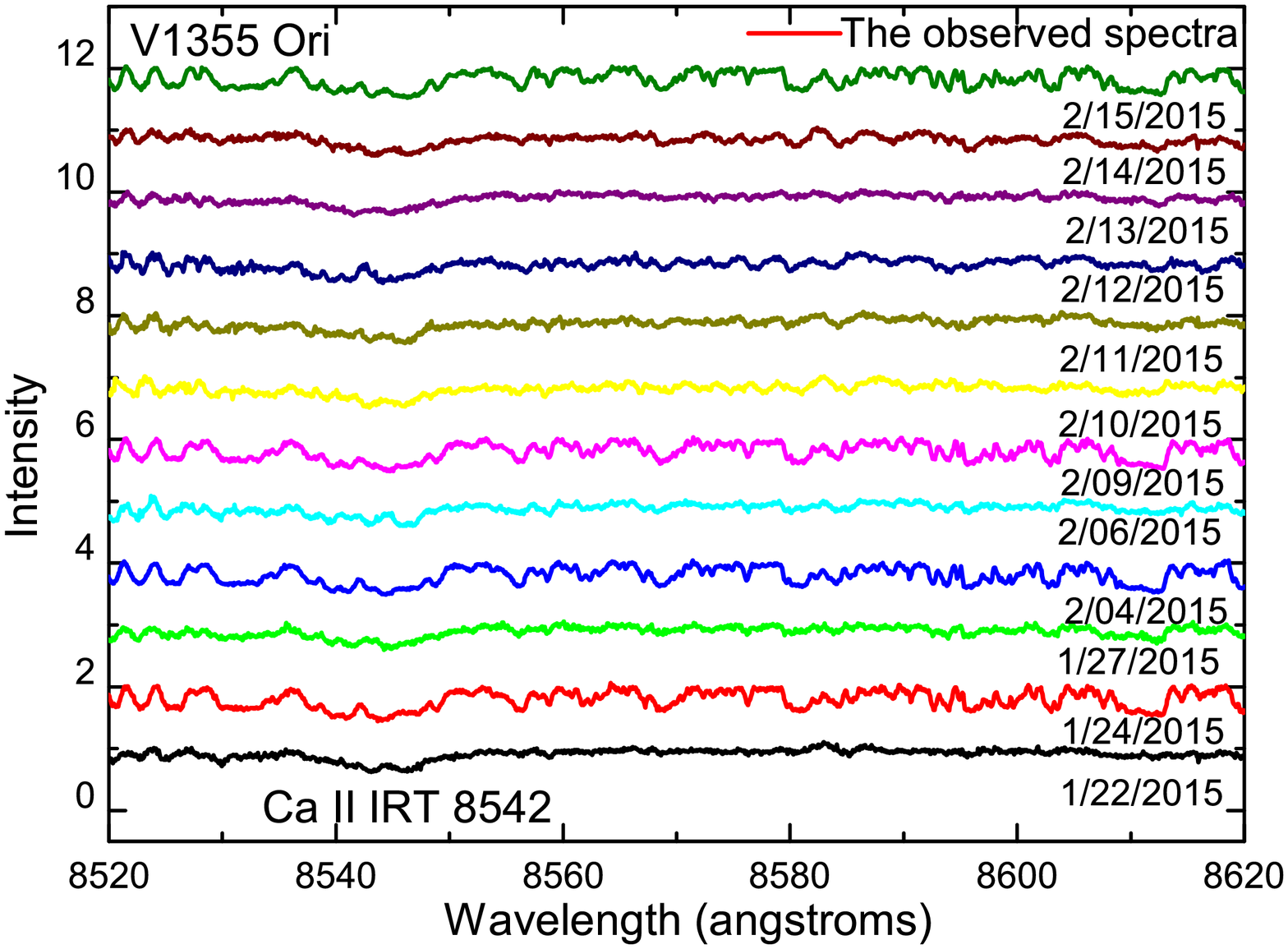}
\caption{All the observed spectra for V1355 Ori using 1.8-m telescope in the $\mbox{He~{\sc i}}$ D$_{3}$, $\mbox{Na~{\sc i}}$ D$_{1}$
D$_{2}$, H$_{\alpha}$, and $\mbox{Ca~{\sc ii}}$ 8542 lines. All these spectra were vertically shifted by 1.}
\end{figure*}

\begin{figure}
\centering
\subfigure[$\mbox{Ca~{\sc ii}}$ 8662 and 8542 lines]{
\includegraphics[width=7.0cm,height=5.7cm]{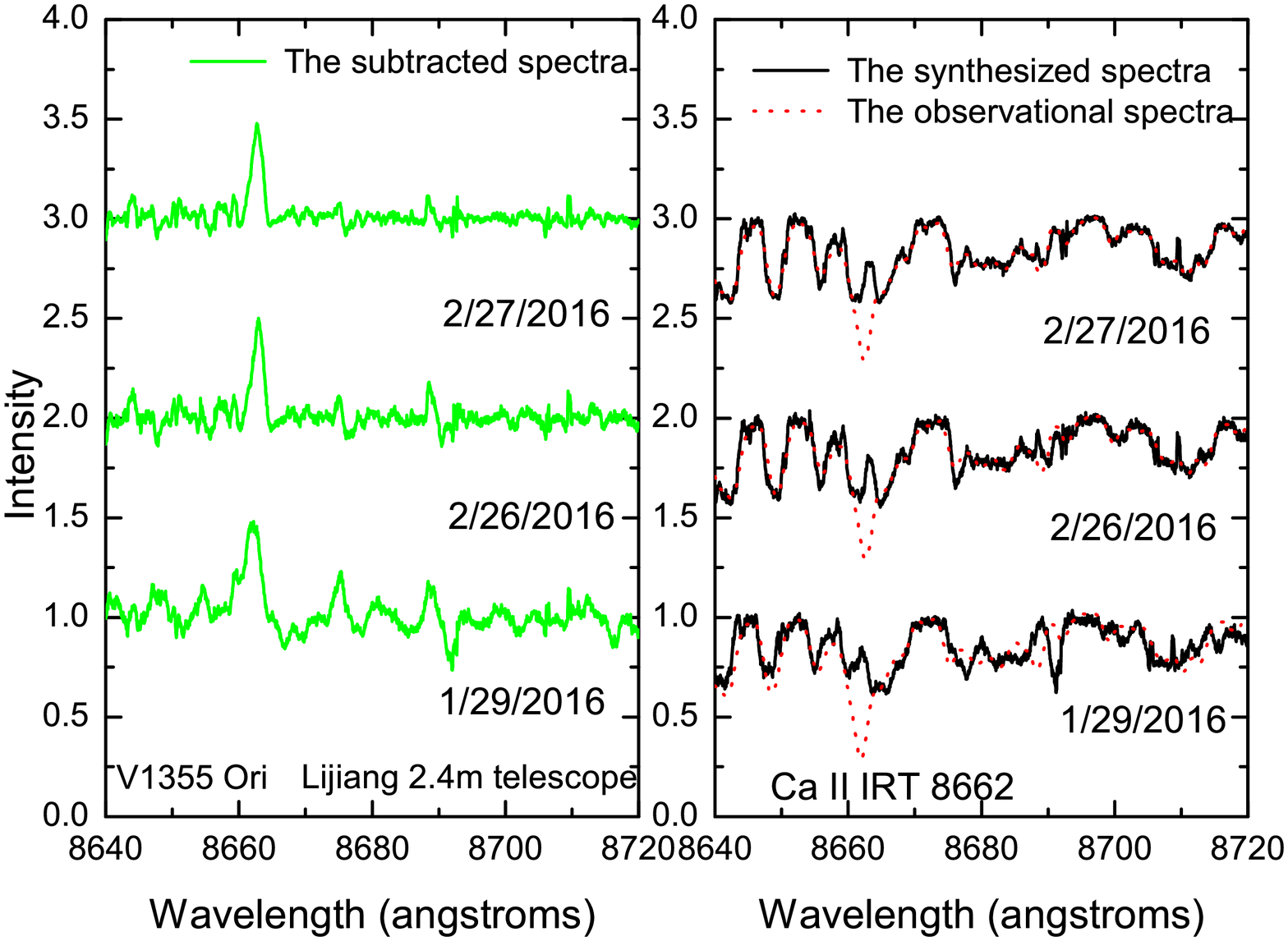}
\includegraphics[width=7.0cm,height=5.7cm]{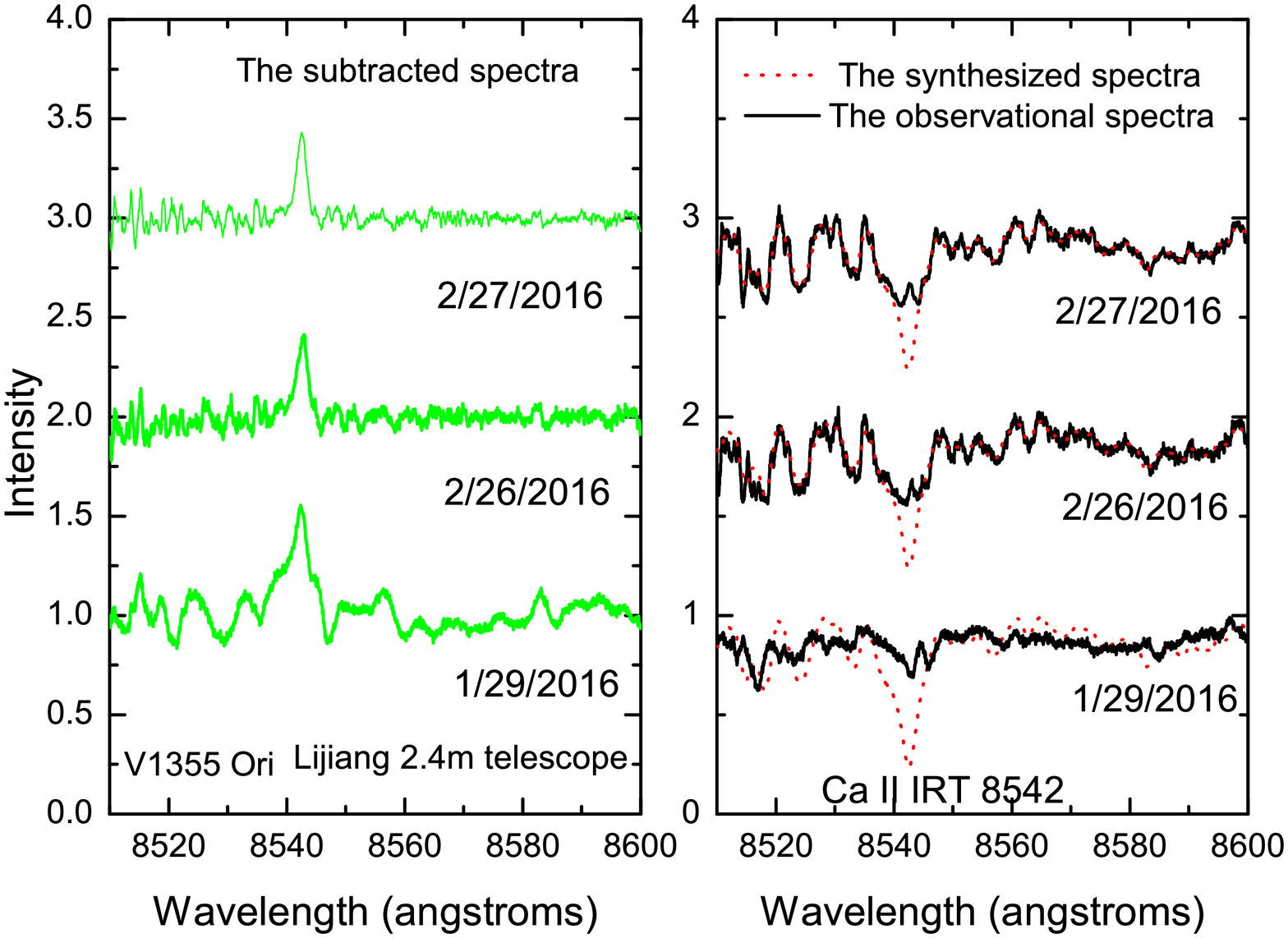}}
\hspace{1in}
\subfigure[$\mbox{Ca~{\sc ii}}$ 8498 and metal lines]{
\includegraphics[width=7.0cm,height=5.7cm]{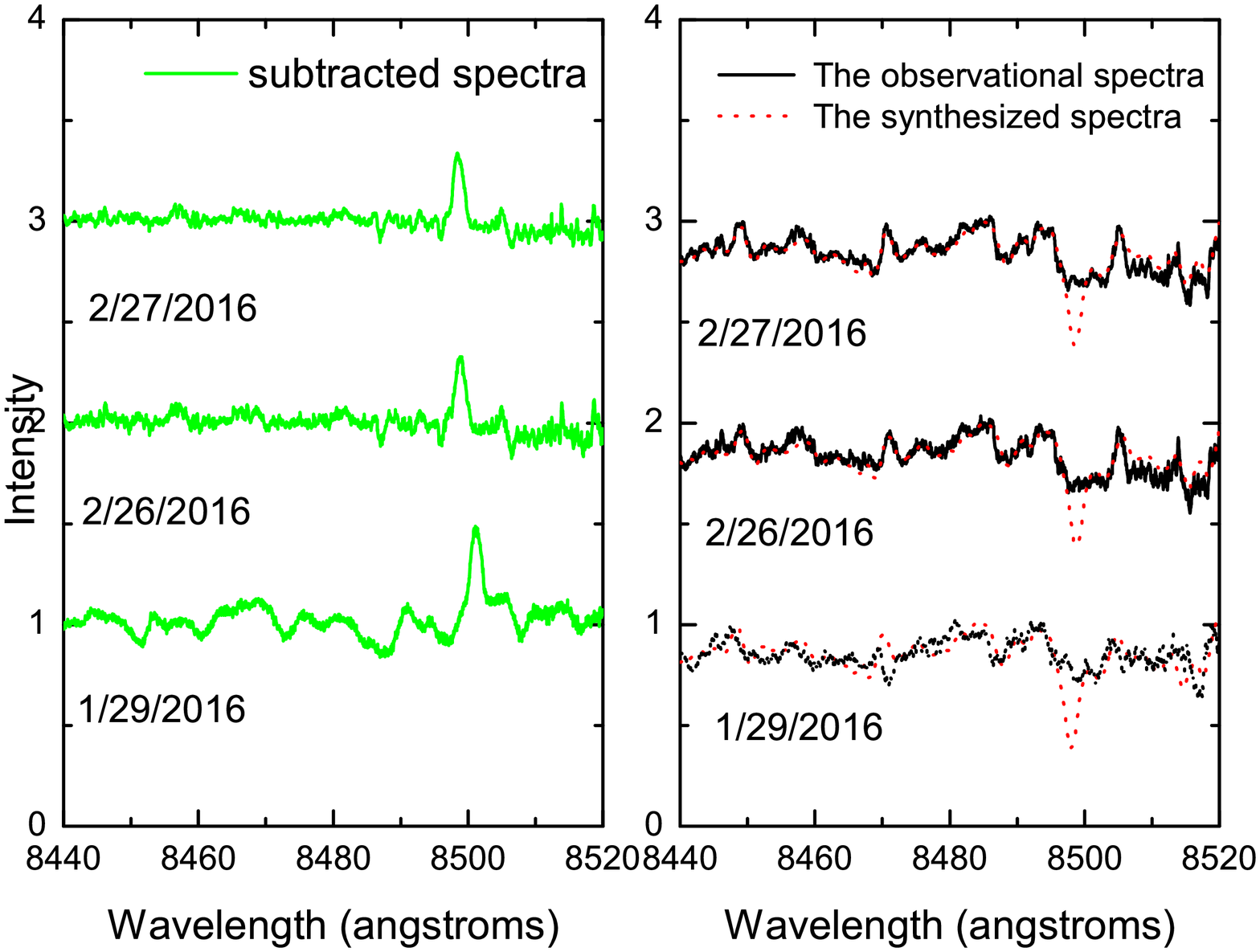}
\includegraphics[width=7.0cm,height=5.7cm]{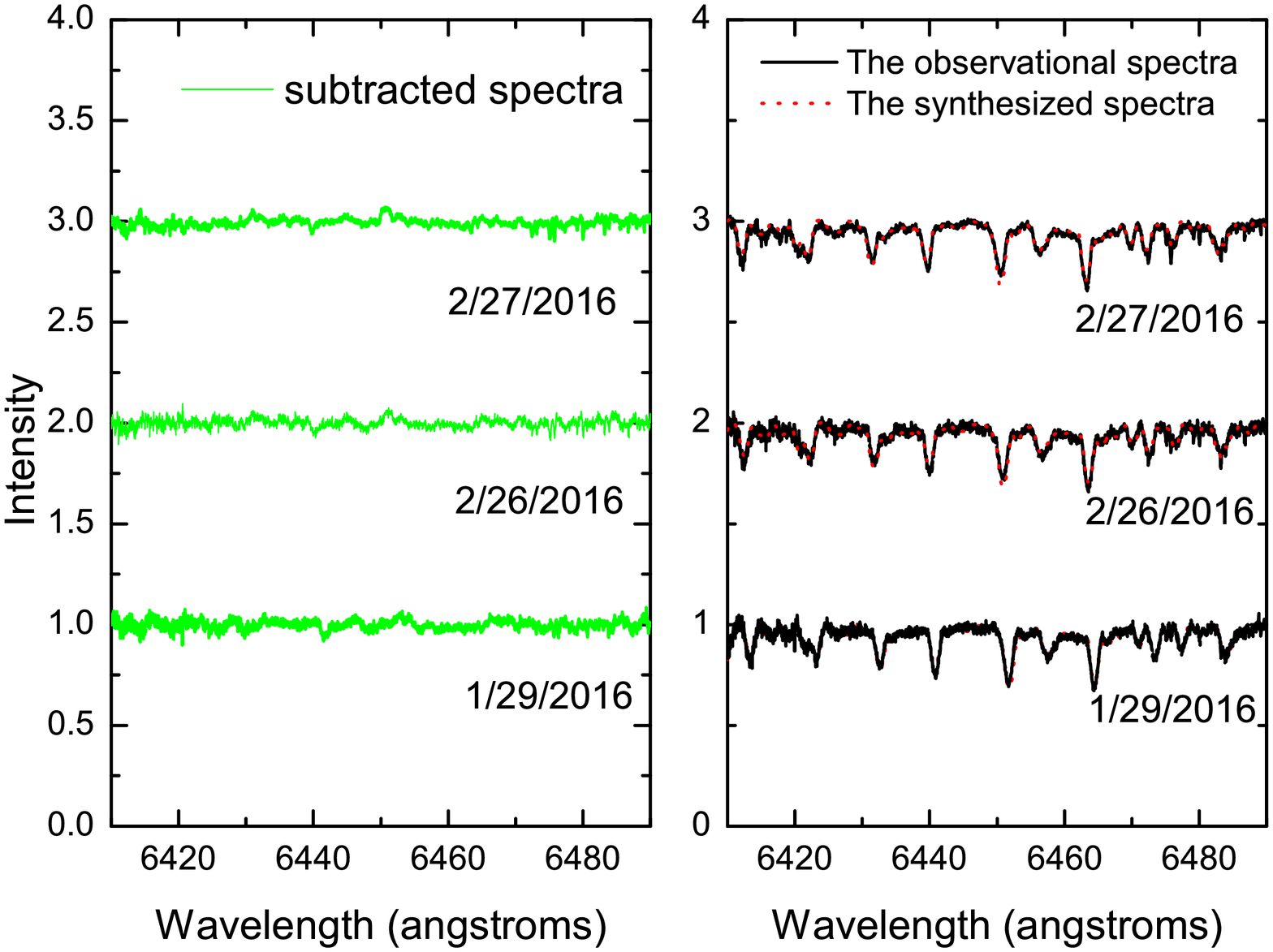}}
\hspace{1in}
\subfigure[H$_{\alpha}$ and $\mbox{Na~{\sc i}}$\ D$_{1}$ D$_{2}$ lines]{
\includegraphics[width=7.0cm,height=5.7cm]{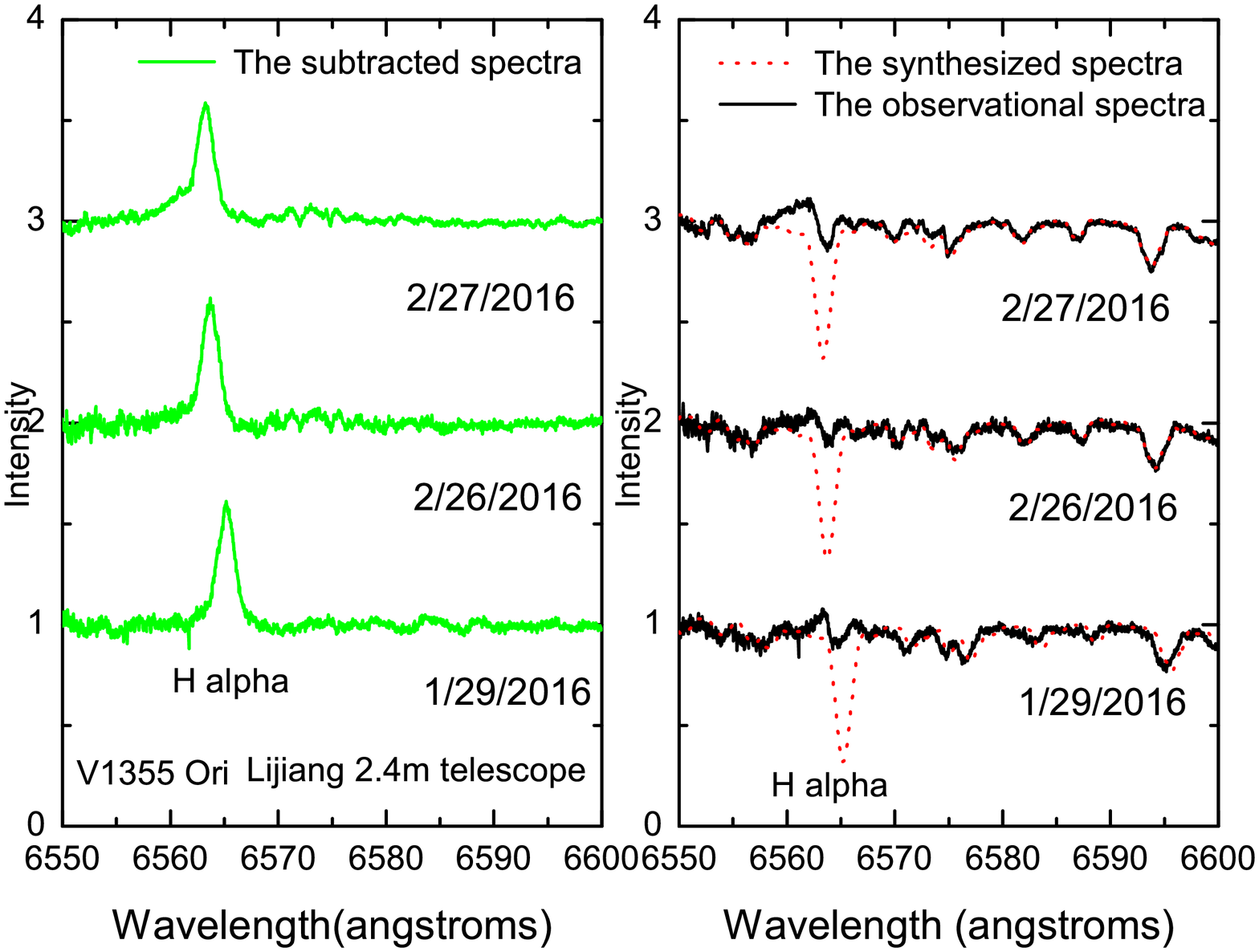}
\includegraphics[width=7.0cm,height=5.7cm]{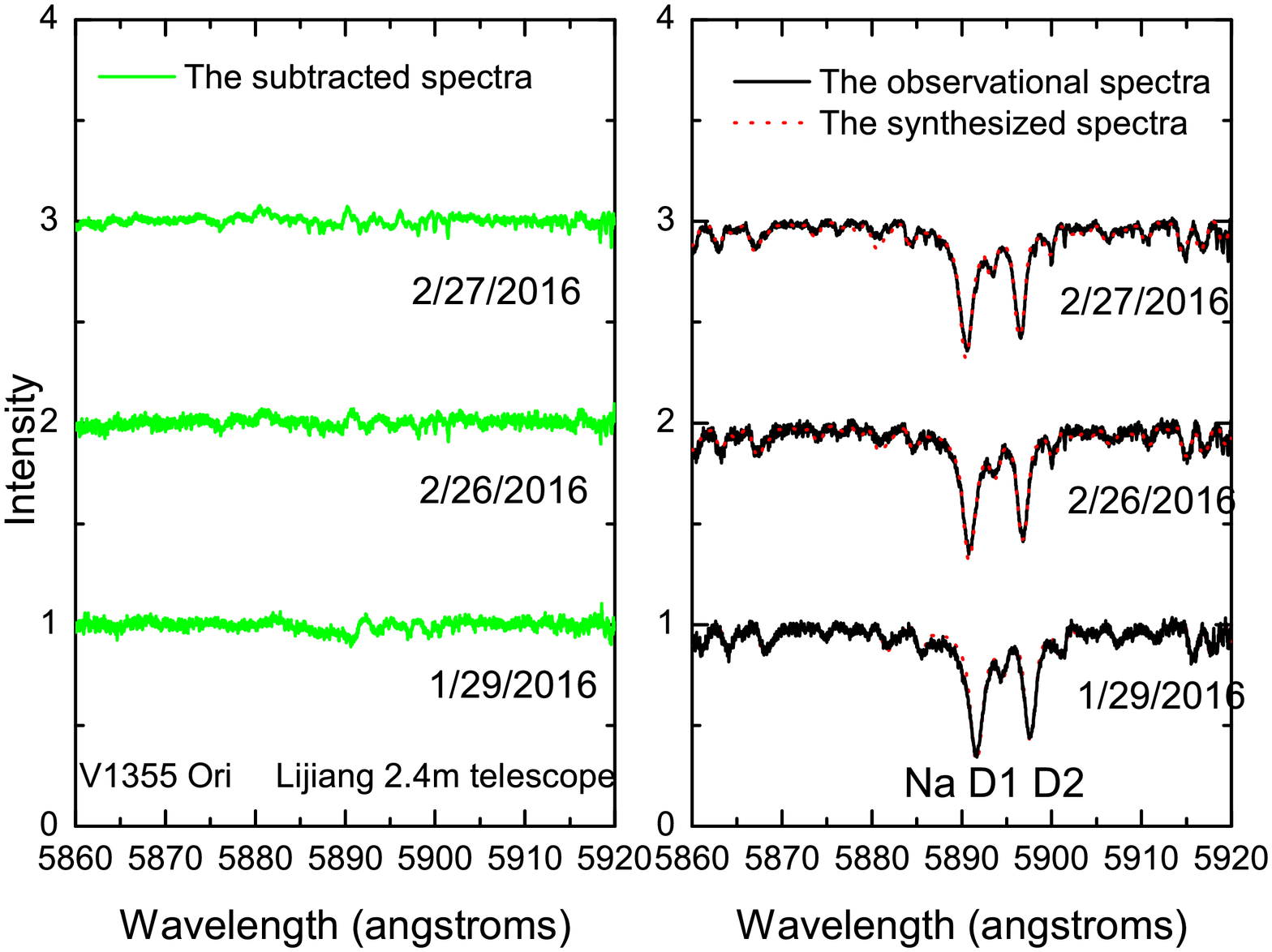}}
\hspace{1in}
\subfigure[H$_{\gamma}$ and H$_{\beta}$, ]{
\includegraphics[width=7.0cm,height=5.7cm]{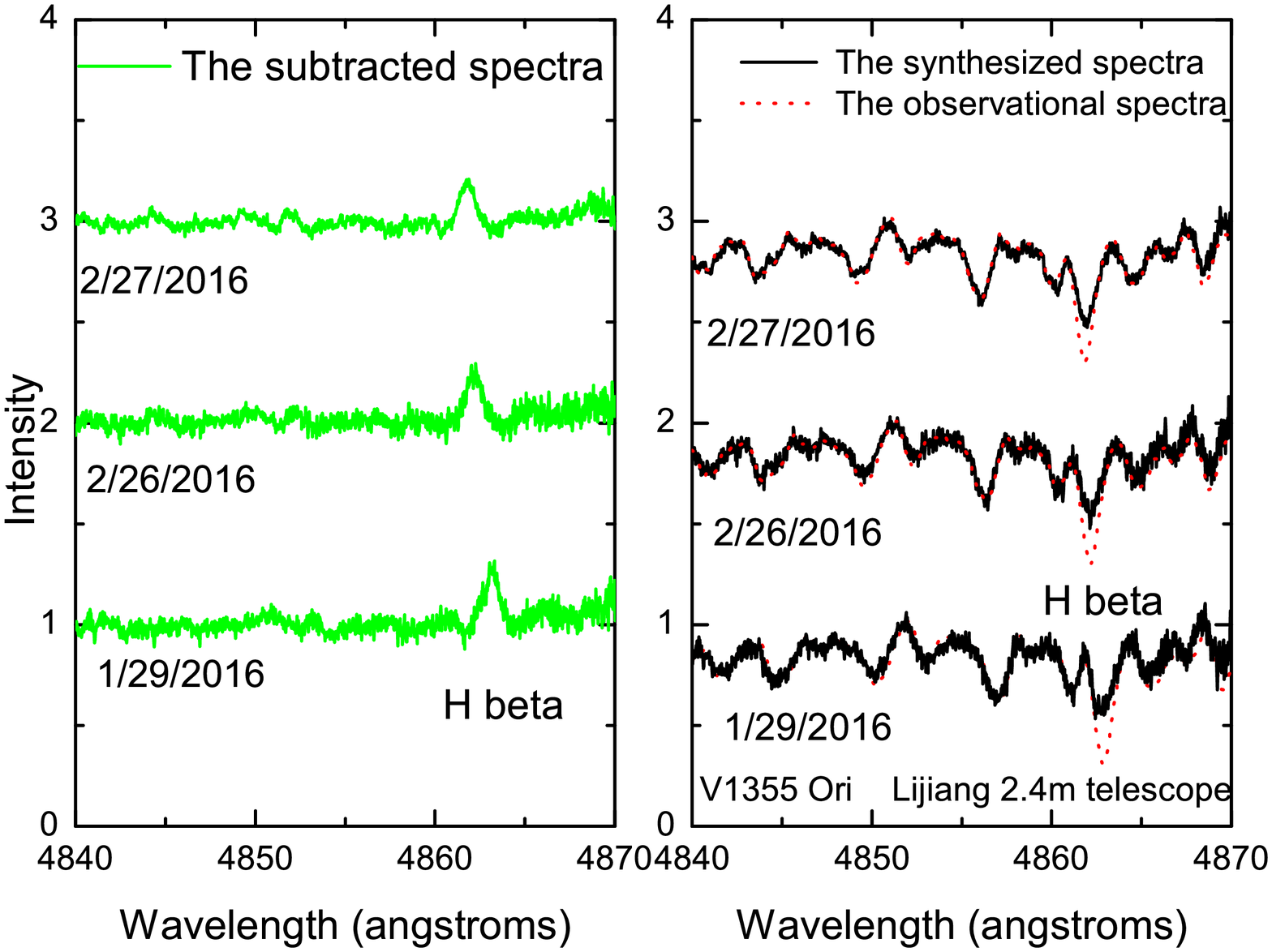}
\includegraphics[width=7.0cm,height=5.7cm]{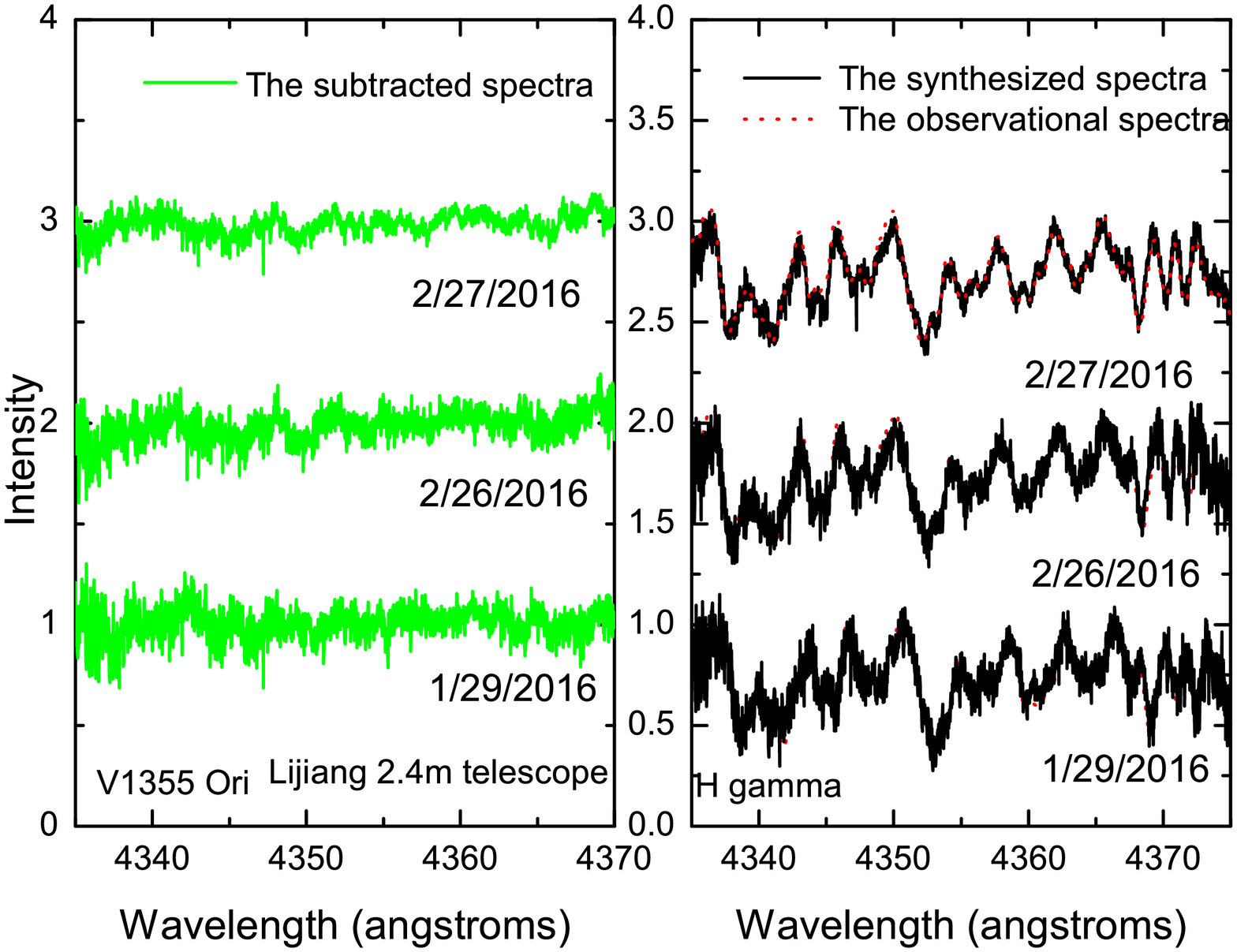}}
\hspace{1in}
\end{figure}
\begin{figure}
\centering
\subfigure[$\mbox{Ca~{\sc ii}}$ H\&K and H$_{\delta}$ lines]{
\includegraphics[width=7.0cm,height=5.7cm]{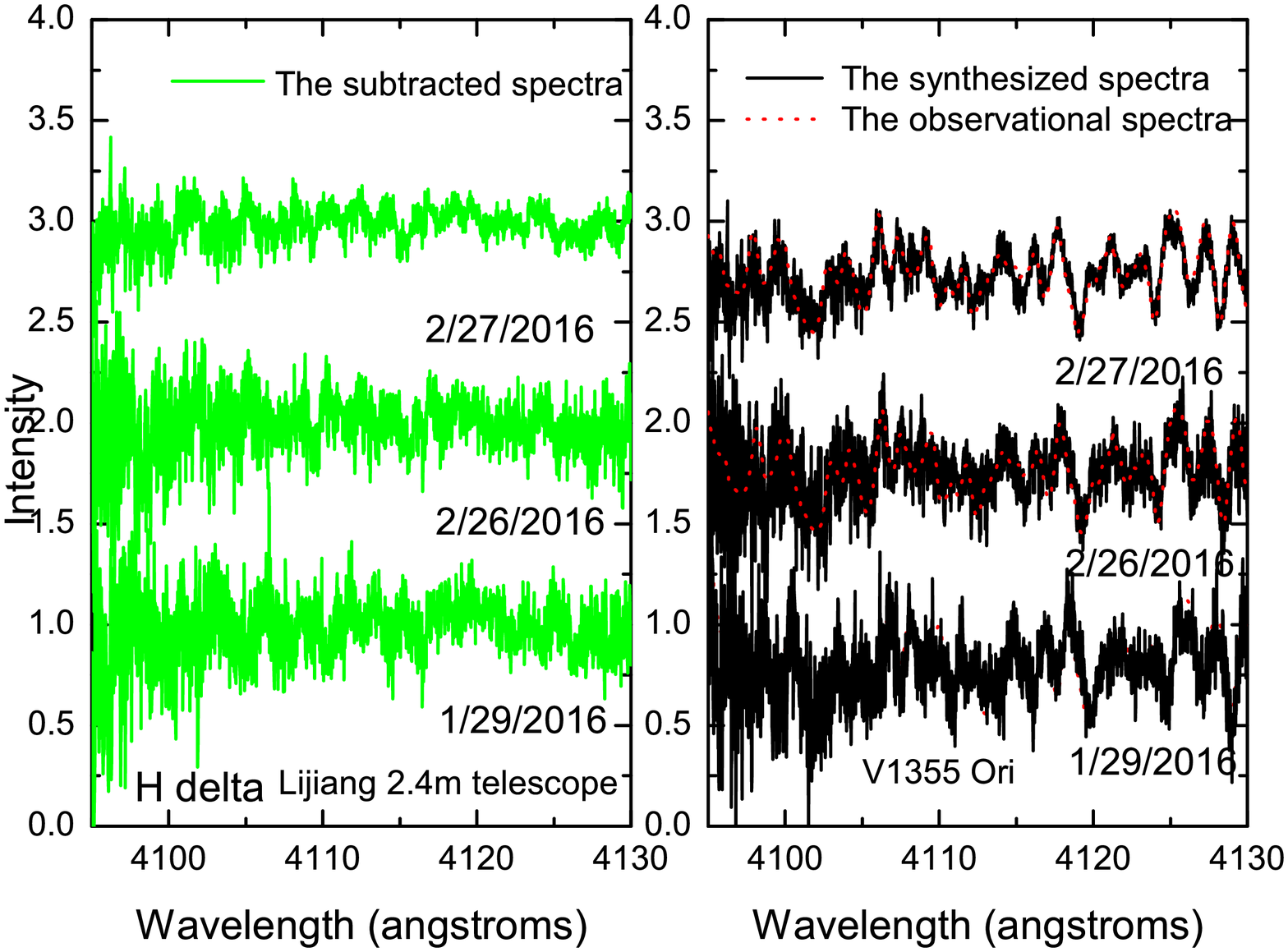}
\includegraphics[width=7.0cm,height=5.7cm]{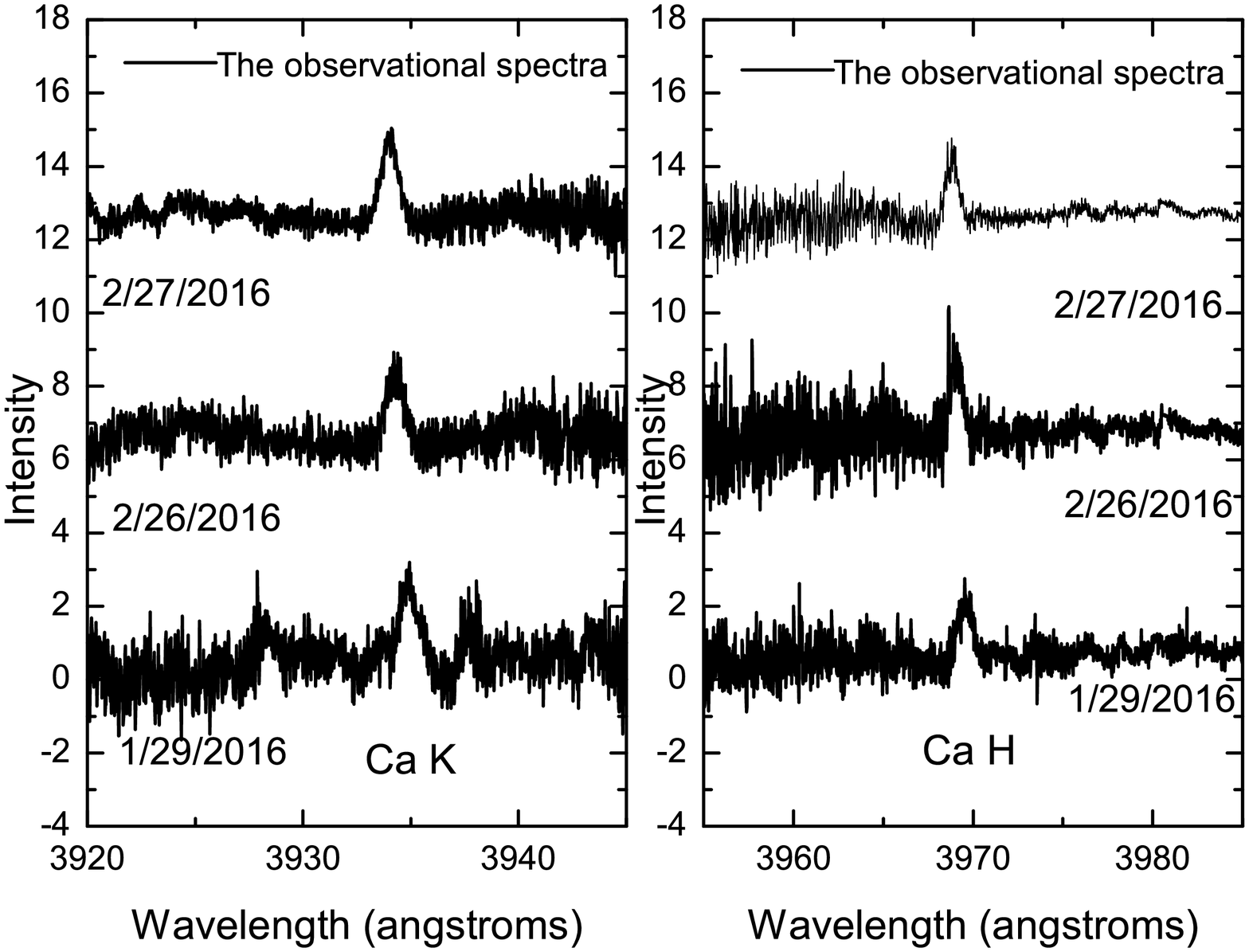}}
\caption{The observed, synthesized, and subtracted spectra of V1355 Ori from the 2.4-m telescope in several spectral lines ($\mbox{Ca~{\sc ii}}$ H\&K, H$_{\delta}$, H$_{\gamma}$, H$_{\beta}$, $\mbox{Na~{\sc i}}$\ D$_{1}$ D$_{2}$, H$_{\alpha}$ and $\mbox{Ca~{\sc ii}}$ IRT lines), and metal lines in the spectral wavelength region of 6400-6500 {\AA}.  All these spectra were vertically shifted by several units.}
\end{figure}

\indent It is of great importance to determine the behaviour of chromospheric active indicators, and chromospheric active properties and evolution (Hall 2008; Montes 2004; Zhang et al. 2015; Zhang et al. 2016a,b; etc). In this paper, we present new high-resolution observations of V1355 Ori. We analysed the properties of chromospheric activity for the $\mbox{Ca~{\sc ii}}$ H\&K, H$_{\delta}$, H$_{\gamma}$, H$_{\beta}$, $\mbox{Na~{\sc i}}$\ D$_{1}$ D$_{2}$, H$_{\alpha}$ and $\mbox{Ca~{\sc ii}}$ IRT lines.\\

\section{Observations}
\indent We obtained new high-resolution spectra for V1355 Ori in twelve nights between Jan. 22 and Feb. 15, 2015 using the Coude Echelle Spectrograph and a 2048$\times$2048 pixels Tektronix CCD detector of the 1.8 meter telescope (Rao et al. 2008; Wei et al. 2000) at the Lijiang station of Yunnan Observatories. The slit width of the spectrograph on the 1.8-m telescope was 61.7-$\mu$ m. The resolution is approximately 50,000 in terms of the FWHM of arc lines. The spectral resolution is approximately 50, 000 and the spectral wavelength region is approximately 5760-11960 ${\AA}$ (Zhang et al. 2016a). The limiting magnitude of the equipment is about 11.5 mag at present. In order to obtain synthesized spectra depicting the stellar photospheric contribution of V1355 Ori, we specifically chose and observed four single inactive stars (HR 617 (K1 III), Beta Gem (K0 III), HR 4182 (G9 IV) and HR 495 (K2 IV)) with spectral types and luminosity classes similar to those of V1355 Ori. HR 617 and Beta Gem were also used as the reference stars for V1355 Ori to discuss the chromospheric activity (Strassmeier 2000; etc).\\
\indent We also observed V1355 Ori using the new high-resolution spectrograph and a 4096$\times$4096 pixels Tektronix CCD detector on the 2.4-m telescope (Fan et al. 2015) at the Lijiang station of Yunnan Observatories in three nights (Jan. 29 and Feb. 26 and 27, 2016). The slit width used in this part of the observation was 62.5 $\mu$ m. The spectral wavelength region is approximately 3890-10600 ${\AA}$. The spectral resolution in terms of the FWHM of the arc comparison lines is 0.110 ${\AA}$ in the $\mbox{Ca~{\sc ii}}$ H\&K lines, 0.145 ${\AA}$ in the H$_{\delta}$, 0.117 ${\AA}$ in the H$_{\gamma}$, 0.152 ${\AA}$ in the H$_{\beta}$, 0.146 in several metal lines (6400-6510${\AA}$), 0.147 ${\AA}$ in the $\mbox{He~{\sc i}}$ D$_{3}$, $\mbox{Na~{\sc i}}$\, D$_{1}$, D$_{2}$ lines, 0.150 ${\AA}$ in the H$_{\alpha}$ line, 0.212 ${\AA}$ in the $\mbox{Ca~{\sc ii}}$ IRT 8498 ${\AA}$ line, 0.245 ${\AA}$ in the $\mbox{Ca~{\sc ii}}$ IRT 8542 ${\AA}$ line, and 0.227 ${\AA}$ in the $\mbox{Ca~{\sc ii}}$ IRT 8662 ${\AA}$ line. The corresponding spectral resolution is about 43, 000 according to the FWHM of the arc comparison lines. The limiting magnitude of the equipment is about 12 mag at present. We also observed HR 617 (K1 III) as a reference star for V1355 Ori to discuss the chromospheric activity.\\
\indent We followed the standard process to reduce the spectra using the Image Reduction and Analysis Facility (IRAF) package, which contains tools to perform CCD image trimming, bias subtraction, flat-field correction, removal of cosmic rays, background subtraction, and multi-spectrum extraction. The wavelength was calibrated using the Th-Ar lamp, and the observed spectra were normalized by a low-order polynomial function. The spectra of V1355 Ori from Lijiang 1.8-m telescope are illustrated in Fig. 1 and the spectra from 2.4-m telescope are plotted in Fig.2. We list our observation log of V1355 Ori in Table 1, which includes observational date, the Heliocentric Julian Date (HJD), exposure times, and signal to noises ratios of the chromospheric active lines.\\
\section{Spectroscopic analysis}
\label{sect:data}
For the observed spectra (Figs. 1 and 2), all the H$_{\alpha}$ spectral lines exhibit variable behaviour (filled-in absorption, partial emission with a center absorption component, or emission above continuum). The $\mbox{Na~{\sc i}}$\ lines demonstrate deep absorptions and the $\mbox{Ca~{\sc ii}}$ IRT lines exhibit filled-in absorptions with minor self-reversal core emissions. The $\mbox{Ca~{\sc ii}}$ H\&K lines show obvious emissions above continuum. The H$_{\delta}$, H$_{\gamma}$, and H$_{\beta}$ lines exhibit absorptions.\\
\indent We analysed the spectra of V1355 Ori using a synthetical spectral subtraction technique (Barden 1985; Montes et al. 1995). The synthesized spectra of V1355 Ori were obtained from rotationally broadened and radial-velocity shifted spectra of a single inactive star with similar spectra type and luminosity class as our object. The rotational velocity value (vsini= 39.6 km~$s^{-1}$) of V1355 Ori was determined using some single metallic spectra lines in the wavelength ranges of 6387-6487 {\AA} (Fig. 2). We chose Beta Gem for the 1.8m telescope and HR 617 for the 2.4m telesccope for this purpose. Our result is similar to the previous results of 40 $\pm$ 0.5 km~$s^{-1}$ derived by Strassmeier (2000) and 46 km~$s^{-1}$ (Osten \& Saar 1998; Strassmeier et al. 1999). All the subtracted, observed, and synthesized spectra are shown in Figs. 2 and 3. The $\mbox{Ca~{\sc ii}}$ IRT lines have a poorer quality than the $\mbox{Na~{\sc i}}$\ and H$_{\alpha}$ lines because of shorter exposure time and poorer weather condition at the time of observation. For $\mbox{Ca~{\sc ii}}$ H\&K lines, we did not used spectral subtraction technique to analyse them because their emission over continuum and low S/N ratio. The subtraction spectra of the $\mbox{Na~{\sc i}}$\, D$_{1}$, D$_{2}$ lines show weak emissions, while all subtracted spectra of the H$_{\alpha}$ line exhibit obvious emission over the continuum. The excess spectra for the $\mbox{Ca~{\sc ii}}$ IRT lines display emissions. There are no excess emission in the H$_{\delta}$, and H$_{\gamma}$ lines, but there is excess emission in the H$_{\beta}$ line. The EWs were calculated by integrating them above the emission lines using the SPLOT package of IRAF. The methods for calculating the EWs and their uncertainties have been published in our previous papers (Zhang \& Gu 2008; Zhang et al. 2014; Zhang et al. 2015; etc). For V1355 Ori, the uncertainties of the EWs are underestimated because the S/Ns for these spectra are low and we do not consider the effect of the error of continuum.\\
\indent The values of the HJD, orbital phase, the excess EWs of chromospheric activity indicators, and the ratios of EW$_{\rm 8542}$/EW$_{8498}$ and E$_{\rm H_{\alpha}}$/E$_{H_{\beta}}$ are listed in Table 2. The orbital phases were calculated using the function: HJD = 2450540.365 + 3.87192*E (Strassmeier 2000). We also plotted the EWs of V1355 Ori against the HJD or phase in Fig. 4, where different symbols represent different chromospheric active indicators.\\
\begin{figure*}
\includegraphics[width=7.45cm,height=6.5cm]{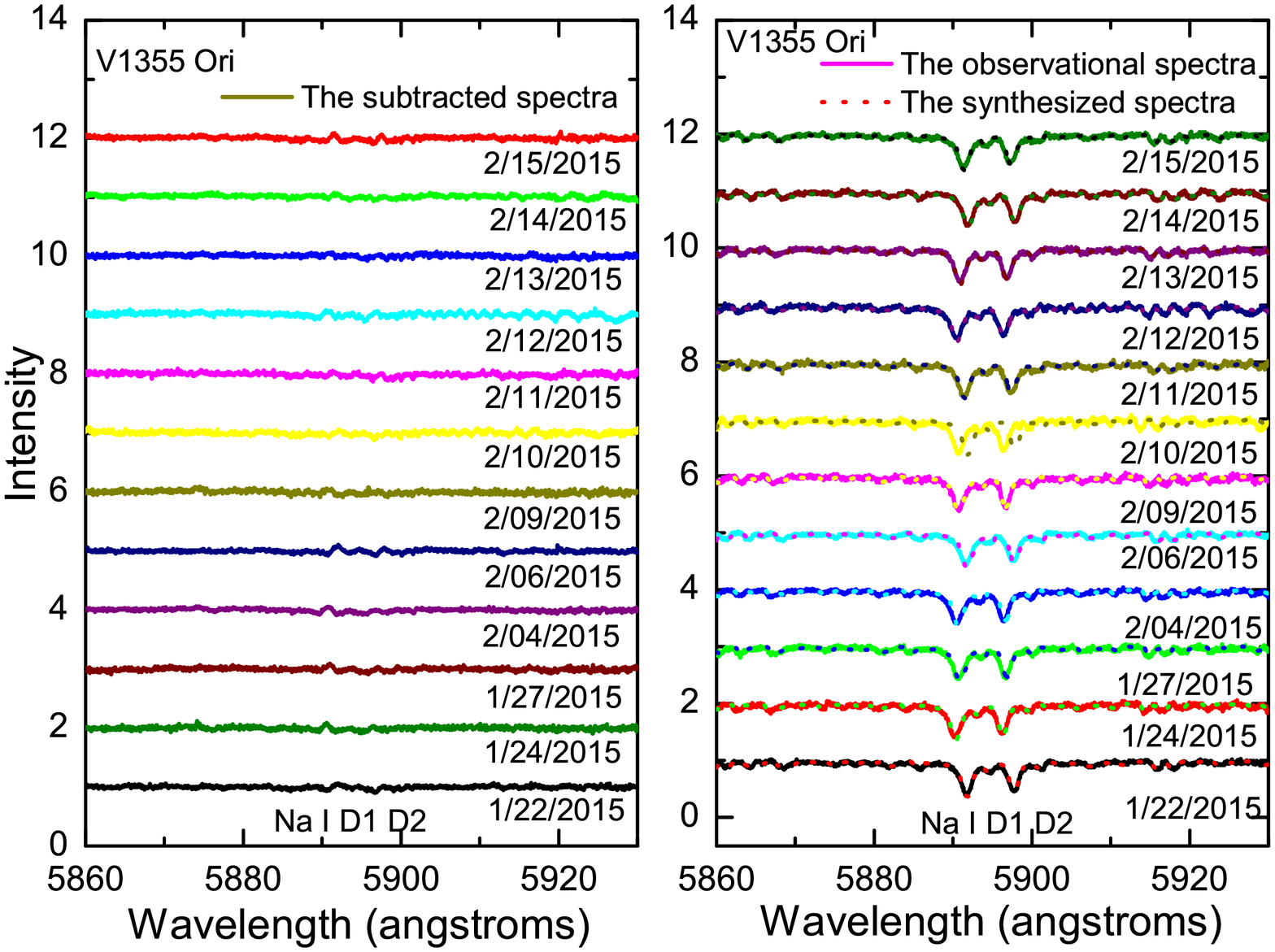}
\includegraphics[width=7.45cm,height=6.5cm]{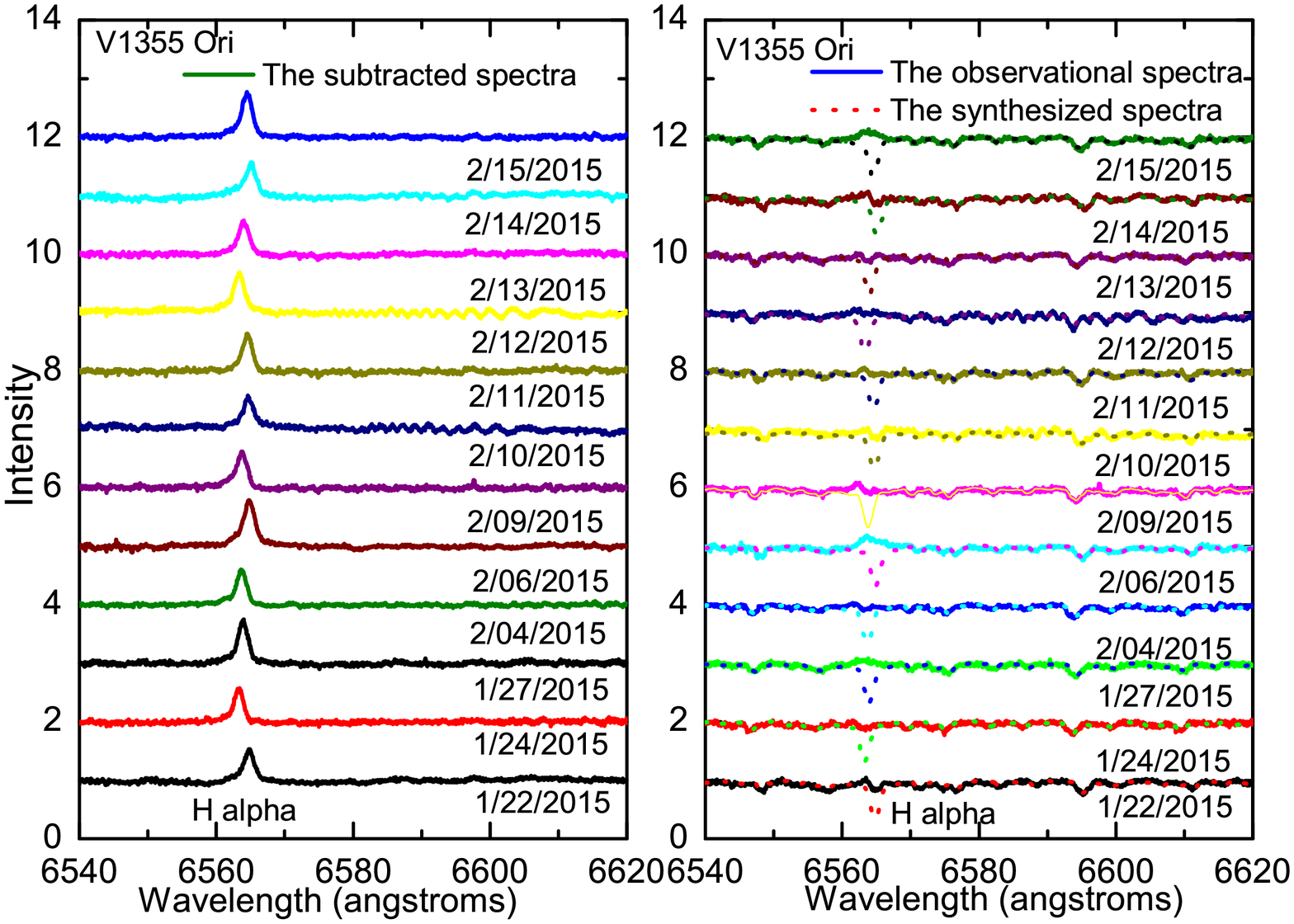}
\includegraphics[width=7.45cm,height=6.5cm]{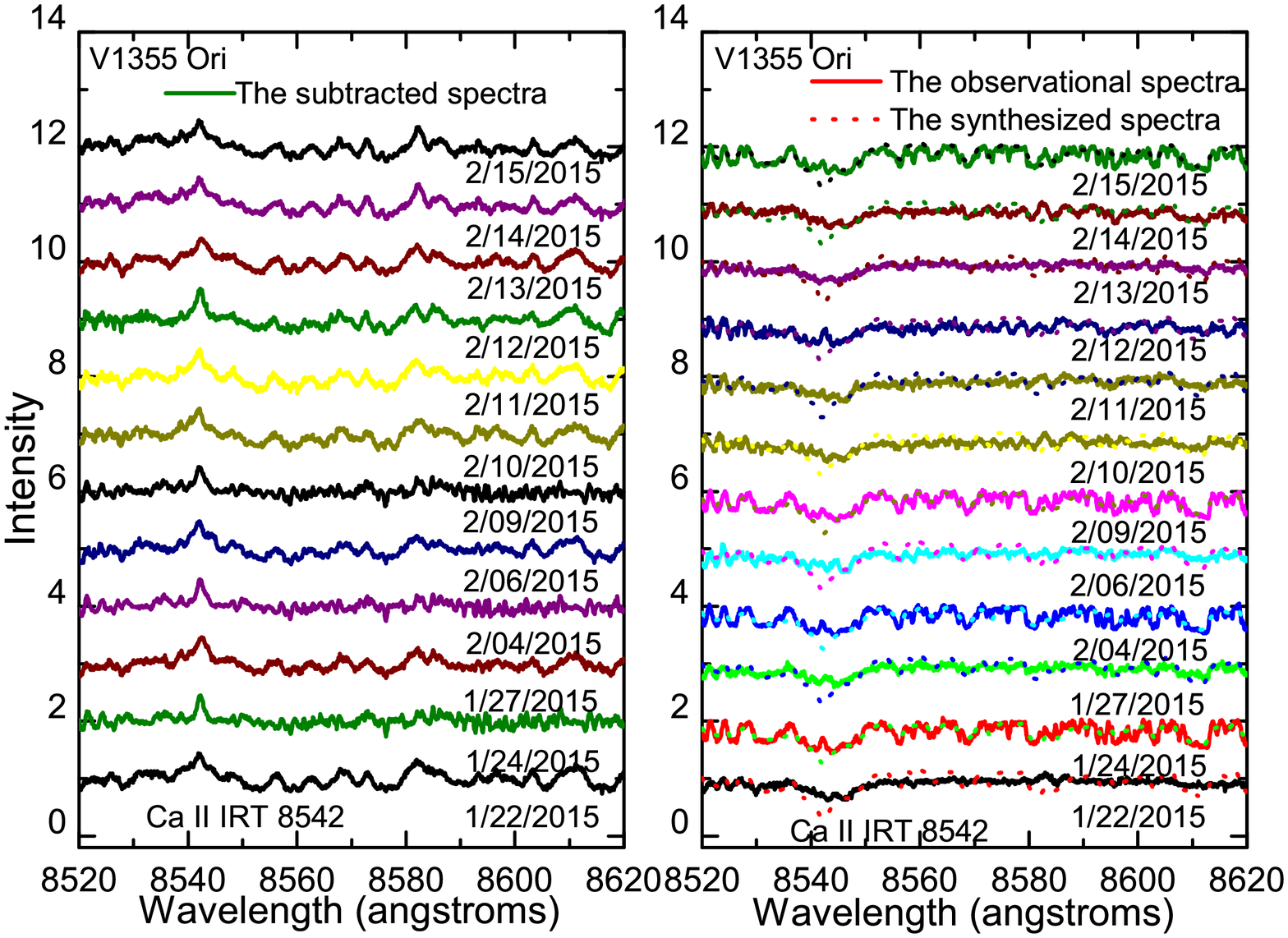}
\caption{Observed and synthesized (left), and subtracted (right) spectra of V1355 Ori using 1.8-m telescope.}
\end{figure*}
\begin{table*}
\tiny\tabcolsep 0.03cm \caption{The values of chromospheric emissions of V1355 Ori}
\begin{tabular}{lccccccccccll}
\hline\hline \multicolumn{1}{c}{HJD(245,)}
&\multicolumn{1}{c}{Phase} &
\multicolumn{1}{c}{EW$_{\rm Ca_{II}K}$(${\AA}$)}&
\multicolumn{1}{c}{EW$_{\rm Ca_{II}H}$(${\AA}$)}&
\multicolumn{1}{c}{EW$_{\rm Na_{I}D_{1}}$(${\AA}$)}&
\multicolumn{1}{c}{EW$_{\rm Na_{I}D_{2}}$(${\AA}$)}&
\multicolumn{1}{c}{EW$_{\rm H_{\alpha}}$(${\AA}$)}&
\multicolumn{1}{c}{EW$_{\rm H_{\beta}}$(${\AA}$)}&
\multicolumn{1}{c}{E$_{\rm H_{\alpha}}$/E$_{H_{\beta}}$}&
\multicolumn{1}{c}{EW$_{\rm Ca_{II}8498}$(${\AA}$)}&
\multicolumn{1}{c}{EW$_{\rm Ca_{II}8542}$(${\AA}$)}&
\multicolumn{1}{c}{EW$_{\rm Ca_{II}8662}$(${\AA}$)}&
\multicolumn{1}{c}{EW$_{\rm 8542}$/EW$_{8498}$}\\
\hline

7045.22327 & 0.009   &        -           &        -        &    0.093  $\pm$    0.011  &  0.055 $\pm$   0.010   &   1.257  $\pm$   0.180&      -            &     -                 & 1.107  $\pm$  0.016    &  1.037  $\pm$   0.189  &  1.31    $\pm$   0.057  & 0.937 $\pm$0.154      \\
7046.23188 & 0.269   &        -           &        -        &    0.087  $\pm$    0.002  &  0.058 $\pm$   0.010   &   1.200  $\pm$   0.016&      -            &     -                 & 0.464  $\pm$  0.029    &  0.646  $\pm$   0.076  &  1.393   $\pm$   0.133  & 1.392 $\pm$0.132      \\
7050.17989 & 0.289   &        -           &        -        &    0.086  $\pm$    0.008  &  0.049 $\pm$   0.001   &   1.663  $\pm$   0.043&      -            &     -                 & 0.967  $\pm$  0.070    &  1.306  $\pm$   0.148  &  0.846   $\pm$   0.086  & 1.351 $\pm$0.113      \\
7058.14409 & 0.346   &        -           &        -        &    0.115  $\pm$    0.009  &  0.057 $\pm$   0.009   &   1.276  $\pm$   0.023&      -            &     -                 & 0.496  $\pm$  0.055    &  0.810  $\pm$   0.106  &  0.959   $\pm$   0.11   & 1.633 $\pm$0.172      \\
7060.11388 & 0.855   &        -           &        -        &    0.137  $\pm$    0.011  &  0.063 $\pm$   0.001   &   2.012  $\pm$   0.098&      -            &     -                 & 1.089  $\pm$  0.087    &  1.295  $\pm$   0.470  &  1.455   $\pm$   0.069  & 1.189 $\pm$0.375      \\
7063.11839 & 0.631   &        -           &        -        &    0.063  $\pm$    0.013  &  0.011 $\pm$   0.005   &   1.371  $\pm$   0.034&      -            &     -                 & 0.614  $\pm$  0.009    &  0.791  $\pm$   0.144  &  0.891   $\pm$   0.039  & 1.288 $\pm$0.226      \\
7064.13065 & 0.892   &        -           &        -        &    0.061  $\pm$    0.002  &  0.016 $\pm$   0.012   &   1.163  $\pm$   0.064&      -            &     -                 & 0.971  $\pm$  0.047    &  0.890  $\pm$   0.338  &  0.945   $\pm$   0.259  & 0.917 $\pm$0.290      \\
7065.09594 & 0.141   &        -           &        -        &    0.043  $\pm$    0.003  &  0.032 $\pm$   0.007   &   1.249  $\pm$   0.025&      -            &     -                 & 1.003  $\pm$  0.094    &  1.028  $\pm$   0.431  &  1.086   $\pm$   0.095  & 1.025 $\pm$0.340      \\
7066.11174 & 0.404   &        -           &        -        &    0.090  $\pm$    0.009  &  0.056 $\pm$   0.005   &   1.425  $\pm$   0.038&      -            &     -                 & 1.039  $\pm$  0.015    &  0.992  $\pm$   0.102  &  1.131   $\pm$   0.149  & 0.954 $\pm$0.082      \\
7067.12605 & 0.666   &        -           &        -        &    0.082  $\pm$    0.008  &  0.017 $\pm$   0.001   &   1.227  $\pm$   0.011&      -            &     -                 & 0.869  $\pm$  0.058    &  1.183  $\pm$   0.262  &  1.113   $\pm$   0.003  & 1.361 $\pm$0.265      \\
7068.12689 & 0.924   &        -           &        -        &    0.086  $\pm$    0.008  &  0.023 $\pm$   0.001   &   1.561  $\pm$   0.034&      -            &     -                 & 0.896  $\pm$  0.014    &  1.028  $\pm$   0.555  &  1.386   $\pm$   0.062  & 1.147 $\pm$0.608      \\
7069.07940 & 0.170   &        -           &        -        &    0.108  $\pm$    0.003  &  0.073 $\pm$   0.005   &   1.765  $\pm$   0.003&      -            &     -                 & 0.544  $\pm$  0.090    &  0.891  $\pm$   0.347  &  1.099   $\pm$   0.228  & 1.638 $\pm$0.576      \\

\hline

7417.15243  & 0.593     &3.992$\pm$0.279      &4.261$\pm$ 0.167&             -             &    -                   &     1.255$\pm$ 0.056  &   0.273$\pm$ 0.010 &5.789	$\pm$ 0.046     & 1.054$\pm$ 0.141       &1.619$\pm$0.091         &1.271$\pm$0.252          &    1.536  $\pm$  0.119  \\
7445.11321  & 0.288     &3.608$\pm$0.092      &2.620$\pm$ 0.230&             -             &    -                   &     1.232$\pm$ 0.030  &   0.267$\pm$ 0.007 &5.811	$\pm$ 0.011     & 0.605$\pm$ 0.024       &0.869$\pm$0.093         &0.926$\pm$0.012          &    1.436  $\pm$  0.097  \\
7446.07349  & 0.536     &4.214$\pm$0.160      &3.834$\pm$ 0.091&             -             &    -                   &     1.546$\pm$ 0.100  &   0.235$\pm$ 0.005 &8.285 $\pm$ 	0.360   & 0.614$\pm$ 0.018       &0.872$\pm$0.038         &0.960$\pm$0.020          &    1.420  $\pm$  0.020  \\

\hline
\end{tabular}
\end{table*}
\section{Discussions and conclusions}
\indent The spectra we observed show obvious emissions above continuum in the $\mbox{Ca~{\sc ii}}$ H\&K lines, and variable behaviour for the H$_{\alpha}$ line. Our results further confirm those behaviour observed in the $\mbox{Ca~{\sc ii}}$ H\&K lines (Strassmeier et al. 2000) and the H$_{\alpha}$ line (Strassmeier 2000; Strassmeier et al. 2000). These spectra show deep absorptions in the $\mbox{Na~{\sc i}}$\ D$_{1}$ D$_{2}$ lines, absorptions for the H$_{\delta}$, H$_{\gamma}$, H$_{\beta}$, and minor self-reversal emissions in the $\mbox{Ca~{\sc ii}}$ IRT absorption lines. These features further demonstrate the chromospheric activity of V1355 Ori. The subtracted spectra exhibit no excess emission in the H$_{\delta}$, and H$_{\gamma}$ lines, trace amount of emissions in the $\mbox{Na~{\sc i}}$ lines, excess emission in the H$_{\beta}$ line, clear emission in the H$_{\alpha}$ line, and weak excess emissions in the $\mbox{Ca~{\sc ii}}$ IRT lines. The maximum EW of our H$_{\alpha}$ line is smaller than that of the large H$_{\alpha}$ flare at HJD 2450909.6 (Strassmeier 2000). \\

\indent The value of the ratio ($EW_{8542}/EW_{8498}$) is also an indicator of the chromospheric activity. For V1355 Ori, the value of $EW_{8542}$/$EW_{8498}$ is in the range of 0.9-1.7. Smaller values indicate optically thick emissions from possible stellar plage event. These values are also consistent with other behaviour of chromospheric activity indicators. Similar values were also discovered in many chromospheric active stars (Cao \& Gu 2005; Ar\'{e}valo \& L\'{a}zaro 1999; Gu et al. 2002; Montes et al. 2001; Zhang et al. 2014, 2015; Zhang et al. 2016a; etc). \\
\indent We calculated the ratio of $E_{H_{\alpha}}$/$E_{H_{\beta}}$ for V1355 Ori in our 2016 run (Table 2). The ratio of excess emission $E_{H_{\alpha}}$/$E_{H_{\beta}}$ was corrected with the function $\frac{E_{H_{\alpha}}}{E_{H_{\beta}}}=\frac{EW_{H_{\alpha}}}{EW_{H_{\beta}}}*0.2444*2.512^{(B-R)}$ from Hall \& Ramsey (1992). The color index of V1355 Ori (B-R)=1.78 were assumed by its spectral type. Buzasi (1989) obtained that low ratios (1-2) can be achieved both in plages and prominences viewed against the disk, but high values (3-15) can only be achieved in prominence-like structures viewed off the stellar limb. We obtained a mean value of 6.628 $\pm$ 0.139 for $E_{H_{\alpha}}$/$E_{H_{\beta}}$ for V1355 Ori. The ratio $E_{H_{\alpha}}$/$E_{H_{\beta}}$ ($\gtrapprox$ 3) implies that the emission of the Balmer lines originates from prominence-like material (Buzasi 1989; Hall \& Ramsey 1992).\\
\indent It is unpractical for us to carry out a detailed study on the chromospheric rotational modulation of V1355 Ori because of data acquisition limitations in the night (we normally only have a partial night) and the total time span required. We had twenty two nights of the 1.8-m telescope for studying V1355 Ori, which corresponds to approximately six orbital periods. Our results do not show any obvious phase modulations in all the chromospheric active indicators lines (Fig. 4). Chromospheric intensity as well as the presence of circumstellar material may dilute or even suppress the rotationally modulated plage signature (Strassmeier 2000). As illustrated in Fig. 4, there is a weak time-variation of excess emission EWs in the $\mbox{Na~{\sc i}}$\, D$_{1}$ D$_{2}$ $\mbox{Ca~{\sc ii}}$ IRT, and H$_{\alpha}$ lines (which is the strongest). The increase of EWs is marked by red arrows in Fig.4. All these indicators are consistent with each other. These phenomena can be explained by plage events, which is consistent with the observed behaviour of these chromospheric active indicators. In the future, further high-resolution spectra will be required to study the chromospheric activity cycle of V1355 Ori, such as those of photospheric cycle reported by Savanov \& Strassmeier 2008.\\

\begin{figure}
\centering
\includegraphics[width=7.2cm,height=5.8cm]{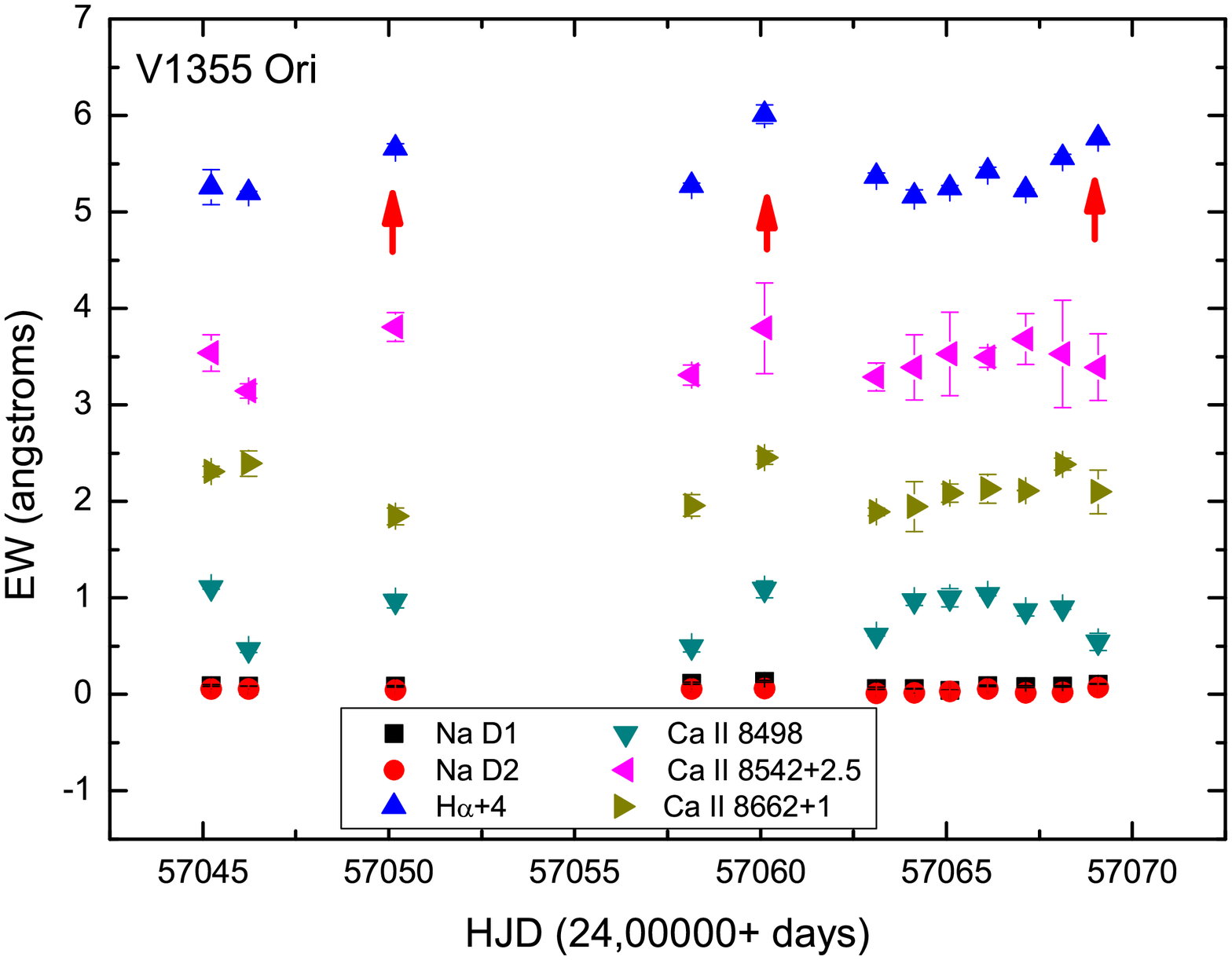}
\includegraphics[width=7.2cm,height=5.8cm]{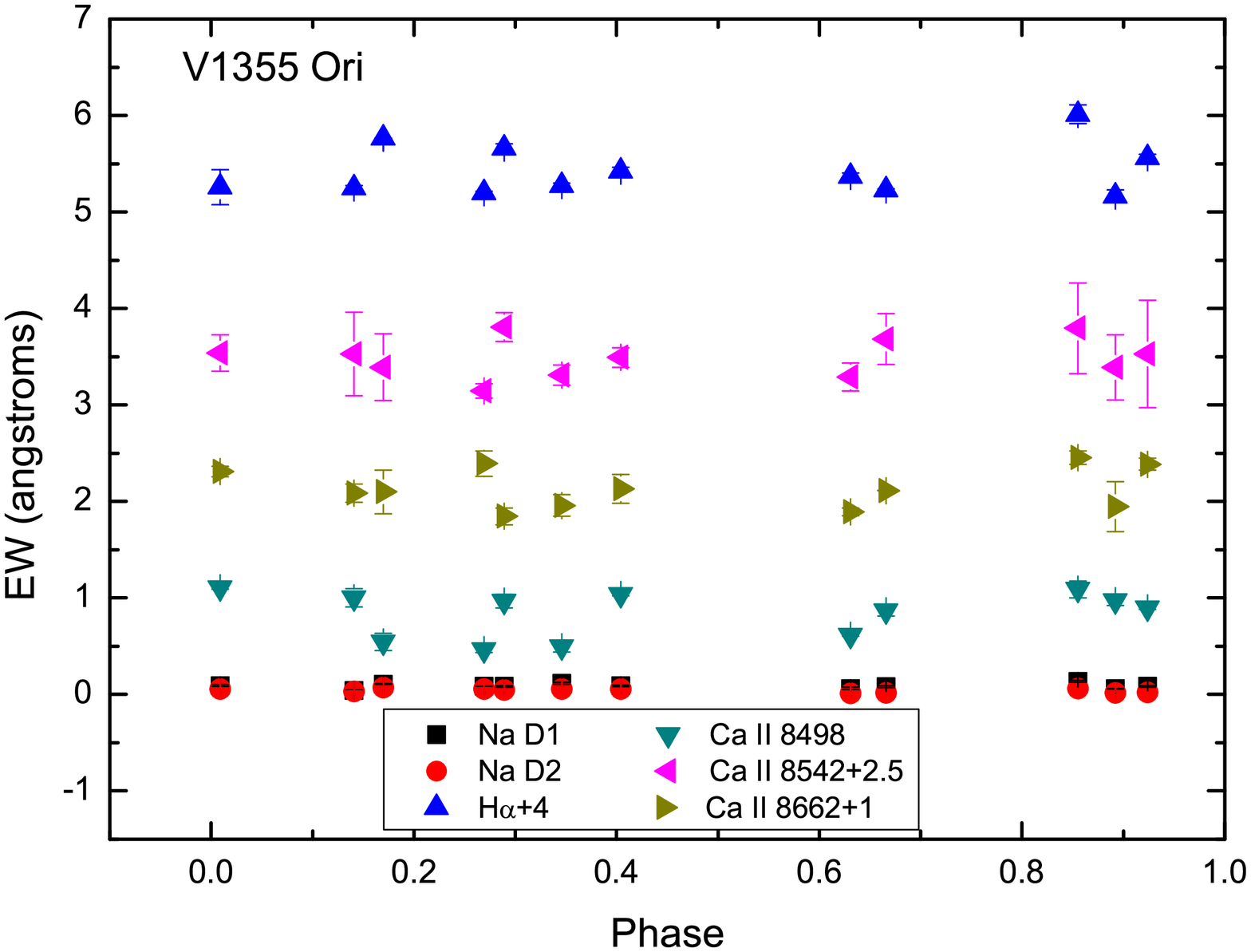}
\caption{EW light curves of chromospheric excess emissions of V1355 Ori vs. HJD (Left) and the orbital phases (right). Red arrows mark the events where EWs increased significantly.}
\end{figure}
\normalem
\begin{acknowledgements}
This work was supported by the Astronomic Joint Fund of NSFC and CAS (U1431114 and 11263001), and science and technology innovation team of Guizhou province (Nos. 20154017). This work was also supported
by the Natural Science Foundation of the GuiZhou Province
office of Education (Grant No. 2014298). We acknowledge the support of the staff of the Lijiang 2.4m
telescope. Funding for the telescope was provided by CAS and the
People's Government of Yunnan Province.
\end{acknowledgements}
\appendix                  

\label{lastpage}

\end{document}